# Observation of confined current ribbon in JET plasmas

E.R. Solano[1], P.J. Lomas[2], B. Alper[2], G. S. Xu[3], Y. Andrew[2], G. Arnoux[2], A. Boboc[2], L. Barrera[1], P. Belo[4], M.N.A. Beurskens[2], M. Brix[2], K. Crombe[5], E. de la Luna[1], S. Devaux[6], T. Eich[6], S. Gerasimov[2], C. Giroud[2], D. Harting[7], D. Howell[2], A. Huber[7], G. Kocsis[8], A. Korotkov[2], A. Lopez-Fraguas[1], M. F. F. Nave[4], E. Rachlew[9], F. Rimini[10], S. Saarelma[2], A. Sirinelli[2], H. Thomsen[11], L. Zabeo[2], D. Zarzoso[12] and JET EFDA contributors[†].

JET-EFDA, Culham Science Centre, Abingdon, OX14 3DB, UK

[1]Laboratorio Nacional de Fusión, Asociación EURATOM-CIEMAT, 28040, Madrid, Spain; [2]Euratom/CCFE Association, Culham Science Centre, Abingdon, Oxon, OX14 3DB, UK; [3]Inst. of Plasma Physics, Chinese Academy of Sciences, Hefei 230031, China; [4]Associação EURATOM/IST, Inst. de Plasmas e Fusão Nuclear, Av Rovisco Pais, 1049-001, Lisbon, Portugal; [5]Department of Applied Physics, Ghent University, Rozier 44, 9000 Gent, Belgium; [6]Max-Planck-Institut für Plasmaphysik, EURATOM-Assoziation, D-85748 Garching, Germany; [7]Forschungszentrum Jülich GmbH, Institut für Plasmaphysik, EURATOM-Assoziation, TEC, D-52425 Jülich, Germany; [8]KFKI, Association EURATOM, P.O.Box 49, H-1525, Budapest, Hungary; [9]Association EURATOM-VR, Department of Physics, SCI, KTH, SE-10691 Stockholm, Sweden; [10]EFDA Close Support Unit, Culham Science Centre, Culham, OX14 3DB, UK; [11]Max-Planck-Institut für Plasmaphysik, EURATOM-Assoziation, D-17491 Greifswald, Germany; [12]Ecole Polytechnique, F-91128, Palaiseau Cedex, France

Abstract: we report the identification of a localised current structure inside the JET plasma. It is a field aligned closed helical ribbon, carrying current in the same direction as the background current profile (co-current), rotating toroidally with the ion velocity (co-rotating). It appears to be located at a flat spot in the plasma pressure profile, at the top of the pedestal. The structure appears spontaneously in low density, high rotation plasmas, and can last up to 1.4 s, a time comparable to a local resistive time. It considerably delays the appearance of the first ELM.

PACS numbers: 52.55 Tn, 52.38 Hb, 47.32 cf

Birkeland first described field-aligned astrophysical plasma current filaments (Birkeland currents) in 1908 [1]. Walén and Alfvén discussed such objects in the 1940's and 50's [2, 3] and experimental confirmation was described in the 60's [4]. More recently, it has been shown that confined (closed field line) field-aligned magnetic vortices are possible stationary solutions of the ideal MHD equations [5] when the density profile is flat. In a stationary plasma the relation $\nabla \times \vec{B} = \mu_0 \vec{j}$ connects a magnetic vortex to a localised current structure, which may itself move with the bulk plasma: this would be a rotating current filament.

---

[†] See Appendix of F. Romanelli et al., Fusion Energy Conference 2008 (Proc. 22nd Int. FEC Geneva, 2008) IAEA, (2008)

Experimental observations of short-lived filamentary current structures in magnetically confined plasmas have been described recently [6,7], and they have been successfully compared to astrophysical current filaments and drift Alfvén vortices [7].

Here we report on a long-lived localised current structure, observed in magnetically confined plasmas at the JET tokamak. They are associated with the MHD fluctuations known as Outer Modes [8, 9, 10] at JET. In this article we identify the Outer Mode (OM) as a confined long-lived rotating current filament or ribbon. Further, the presence of this localised current structure substantially alters plasma behaviour: a quasi-stationary state is established in which sudden energy bursts (Edge Localised Modes, ELMs) are suppressed. Usually in the high confinement regime (H-mode) a region of high gradients of density and temperature is formed, known as the pedestal. The pedestal is eroded quasi-periodically by the spontaneous occurrence of ELMs. Reliable high confinement without ELMs would greatly simplify the design of tokamak fusion reactors.

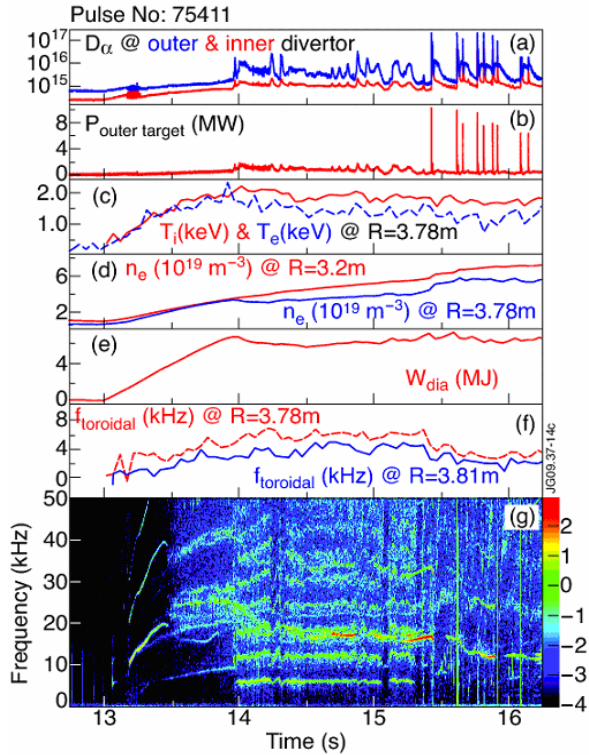

Fig. 1: time traces showing a long-lived Outer Mode, starting from 14 s to 15.38s, and its effect on plasma behaviour: a) rise in $D_\alpha$ ; b) rise in power to outer target; c) drop in pedestal temperatures; d) slower pedestal density rise, rising core density; e) stationary energy; f) toroidal rotation frequency at top and middle of pedestal g) FFT of magnetic signal

Let us first describe the Outer Mode. A very characteristic signature is shown in the fast Fourier transform (FFT) of the Mirnov signals (magnetic probes measuring the poloidal field time derivative): its harmonic structure, shown in Fig 1.g (other plots in that figure will be described later). Typically the fundamental harmonic frequency of the OM is f~ 5-10 kHz and the toroidal mode number is n=1. Every subsequent harmonic has n increased by 1. In the pulse depicted in Fig. 1 the fundamental frequency is f~ 6 kHz and harmonics are seen up to 45 kHz (n=7), but the mode is sometimes clearly recognisable up to 90 kHz (n=15). This rich harmonic structure hints at strong localisation of the current source that produces the magnetic fluctuation. It is observed in all Mirnov probes around the plasma.

The raw magnetic signals contain more information than their spectrum. Plotted in Fig. 2 are signals from a toroidal array of Mirnov coils, measuring the tangential field at different toroidal angles around the device. Plotted at the bottom of the same figure is an integrated signal: discrete blips can be clearly seen, propagating toroidally with f= 6 kHz, due to a localised current structure. This current structure rotates in the same direction as the ions in the plasma and carries excess current relative to the axisymmetric equilibrium current profile. Since the current feature is long-lived (> 1 s), it must be located inside the magnetic separatrix. Since n=1 and the fundamental frequency is narrow, it must be a closed field line, aligned with the magnetic field at a rational surface. To describe the topology of closed field lines it is convenient to use the safety factor q= m/n, where m is the number of poloidal turns and n the number of toroidal turns required for the line to close on itself. Plotted in Fig. 3 is a comparison of measured magnetic signals at various locations around the vessel with simulated signals produced by assuming a field-aligned current ribbon located at the q=4, 3 and 2 surfaces. The current in the ribbon necessary to reproduce the measured signal height is quoted in each box. The scatter in simulated current values is representative of uncertainties in reconstruction of the plasma equilibrium. Comparison of the shapes and phases of measured and simulated signals shows that the current structure is best simulated as a current ribbon with excess current of 100-300 A (plasma current is 2.5 MA), with toroidal mode number n=1, poloidal mode number m=4 (q=4), and toroidal width of the order 5-10% of the toroidal circumference of the plasma, rotating toroidally at 6 kHz. The width of the ribbon is determined by the fastest observed change in dB/dt. Data in one inboard coil is substantially better fitted with q= 4, rather than by q= 2. A q=3 current structure cannot fit the inboard data. Outboard coils, although more numerous (not shown) provide insufficient information to select m. A representation of the q=4 current ribbon that best matches the data is shown in Fig. 4.

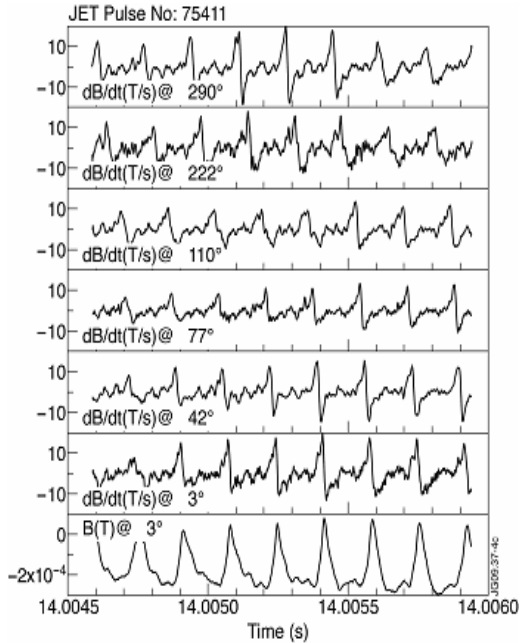

Fig. 2: Mirnov coil signals from toroidal array (outboard), showing toroidal propagation of magnetic feature in co-current direction. The lowest plot is the integrated signal from the coil at 3°

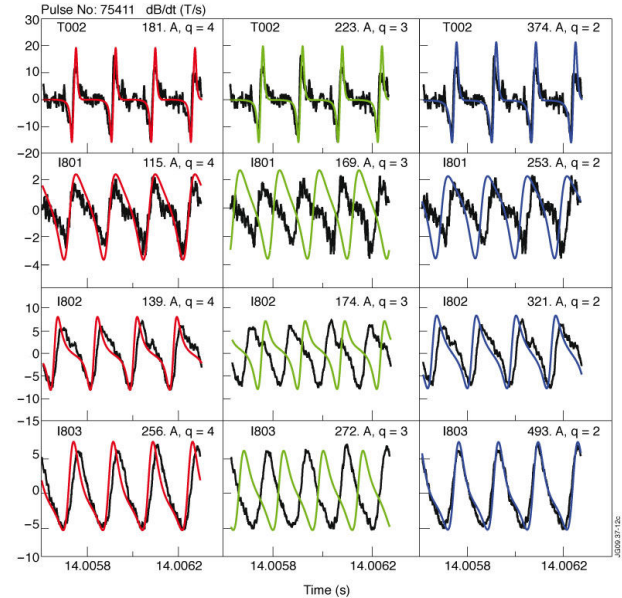

Fig. 3: Comparison of simulated signals with data (black) in Mirnov coils located at the plasma outboard (top row), and three inboard coils located above, near and below the plasma midplane, respectively. Simulated signals are generated with rotating current ribbons located at q=4 (red), q=3(blue) and q=2 (green) flux surfaces. The currents quoted in each case are adjusted to match calibrated signal amplitude.

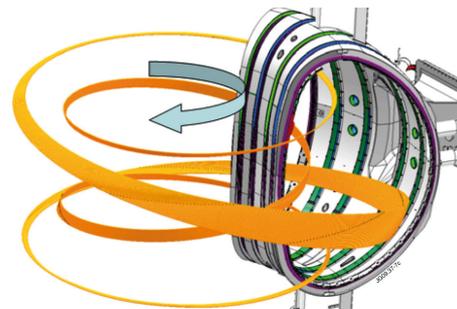

Fig.4: q=4 current ribbon that provides best fit to magnetic signals from Mirnov coil set, with current 100-200 A.

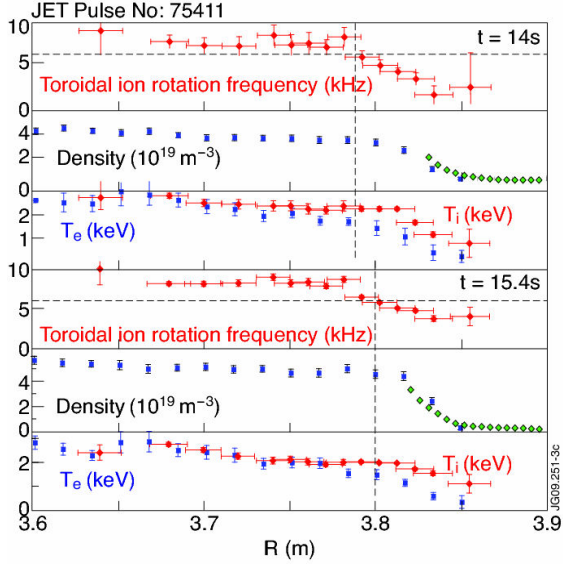

Fig. 5: profiles of toroidal rotation frequency, $n_e$, $T_e$, $T_i$ at start (top 3 plots) and end (lower 3 plots) of OM. Vertical line marks position where mode and plasma rotation frequencies are equal.

Characteristic profiles at the onset (t=14 s) of the OM phase are shown in Fig. 5. From the charge-exchange measurements of the toroidal ion rotation frequency and in the usual assumption that the fluid velocity is dominated by the ion velocity, we can identify the radial position of the current ribbon as the location where toroidal rotation frequency matches mode frequency. From the start of the OM it appears that the location of the mode is at the edge of the ion rotation gradient region, at the flat-top of the density pedestal, inboard of the maximum pressure gradient. Because the rotational shear is high around 6 kHz, the current structure must be well localised in the radial direction. At the end of the OM the toroidal rotation shear is somewhat eroded and the mode location shifts outwards by 2 cm. Nevertheless, the mode is still located in the toroidal rotation gradient region, inboard of the density and temperature high gradient regions. After the 1st ELM the pedestal rotation frequency drops below 4 kHz.

The spectra of several fluctuation measurements show multiple harmonics of the OM near the gradient region. Fluctuations of electron cyclotron emission in the $T_e$ gradient region, X-mode reflectometry and edge channels of Soft X-Rays, all show at least 3 harmonics of the OM. All these fluctuations measurements are sensitive to the flux surface deformation introduced by the current structure. The data is compatible with earlier studies [9] that established the non-tearing character of the OM: a tearing mode would exhibit alternating flat and steep profiles as the magnetic island moves past the measuring channels. These are not observed. It is not presently possible to establish from these fluctuation measurements if the current ribbon is located at the top of the pedestal. It is the high resolution rotation profile, as shown in Fig. 5, that allows us to identify the location of the current source at the pedestal top.

Let us now consider the conditions necessary for the OM to appear, and its impact on plasma behaviour. The JET experiments in which these localised current structures were observed were originally intended to study the stability of plasmas with high temperature pedestals. High temperature pedestals can be obtained in JET, transiently, by reducing particle fuelling (from external sources and wall) and operating at initially low density in the hot ion H-mode regime [11]. The transition to H-mode occurs at low plasma density, and after it the pedestal density, rotation and temperatures rise up until the occurrence of the first natural ELM. Recent examples at JET reach $T_{e,ped}$= 2.8 keV before the first ELM. Obstacles to reach higher pedestal temperatures are washboard modes [12,13], the OM and/or the first large ELM. The plasmas described in this letter had high triangularity ($\delta$=0.4), 2.5 MA of plasma current, 2.7 T toroidal field, and ~15 MW of Neutral Beam Injection heating.

Time traces of an especially long OM were shown in Fig 1. The profile quantities are taken at R=3.78 m, the top of the pedestal, inboard of the pedestal knee, and at the plasma core. The spectrum (FFT) of a Mirnov coil was already discussed. After the transition to H-mode, marked by the initial $D_\alpha$ drop at t=13.2 s, the pedestal rises as usual in an H-mode, until the OM begins at the time t= 13.98 s. Then $D_\alpha$ rises, the pedestal $T_e$ begins to drop, the toroidal rotation frequency remains fairly constant. Density continues to rise, albeit slower than before. The H factor (describes normalised energy confinement, not shown) drops down from a maximum value of $H_{98}$=1.4 to a steady value of $H_{98}$=1.1 at the outset of the OM. A value of $H_{98}$~1 is typical in ELMy H-modes. Impurity content during the OM is normal compared to a medium density ELMy H-mode, as measured by two diagnostics: the line-averaged $Z_{eff}$ from visible bremsstrahlung deconvolution is of order 2, and the charge-exchange based measurement results in $Z_{eff}$~1.5. The OM survives 2 sawtooth crashes (at 14.23 and 14.87 s), indicating that it is a robust feature of the plasma. There is a brief quiet interval between the end of the OM at 15.375s and the ELM at 15.430 s. In this particular pulse the OM does not return after the first ELM. In other pulses OMs sometimes return with ever decreasing duration some hundreds of ms after the first few ELMs, and are typically terminated by ELMs.

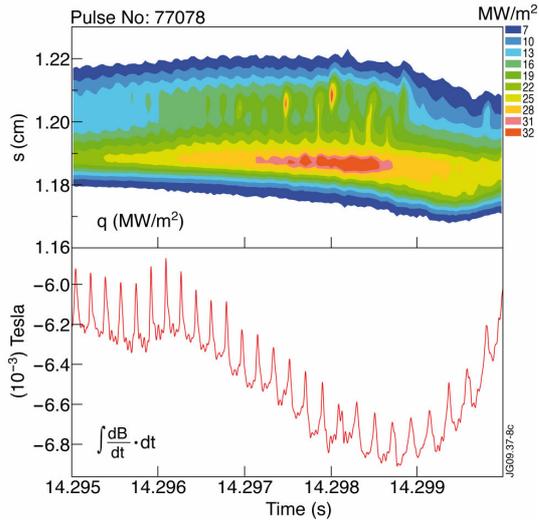

*Fig. 6: heat flux contours at outer strike, measured with IR, and integrated magnetic signal showing correlation of heat pulses with rotation of current structure.*

Comparison with similar pulses allows us to establish that the ELM-free period is prolonged by the presence of the OM [14]. The OM often appears when recycling and gas fuelling are particularly low, and it can be reduced or eliminated by increased fuelling. For example, a recent pulse with the same conditions as in Fig. 1 but with more fuelling (1.9 $10^{21}$ electron/s instead of 1.5 $10^{21}$ e/s) had no long-lived OMs and the first ELM occurred 1 s sooner, at 14.4 s.

Note that the resistive current diffusion time for a 1 cm wide current ribbon at the pedestal top is of order 0.8 s, and in the example shown in Fig. 1 the OM lasts almost 1.4 s. A mechanism is necessary for the mode to survive against resistive decay.

The power outflux to the outer divertor target was measured with a fast Infra-Red (IR) camera, viewing a narrow toroidal strip across the inner and outer target. In the pulse shown in Fig. 1 the time averaged heat flux during the OM is very similar in peak value and radial distribution to the ELM-averaged heat flux, albeit after ELMs there are large strike point position shifts [15]. But the frame rate was 11 kHz, too slow to detect the 6 kHz bursts.

For a pulse with a transient OM (JET 77078, OM from 14.28 to 14.355 s), the camera frame rate was especially set to 1/38 μs (26 kHz), with an integration time of 27 μs, viewing only the outboard target. In this pulse soon after the mode onset the total power to the target increases, consistently with the loss of core confinement at that time. More interestingly, periodic bursts of heat arrive away from the maximum deposition location, as shown in Fig. 6. Both, overall transport across the pedestal and a time-localised (presumably toroidally localised) "hose-pipe" contribute to the increased outer target heat flux during the OM, relative to the previous high confinement phase. The heat deposition, as observed both in this pulse and in the steady OM case, is consistent with the effect of a rotating current structure at the top of the pedestal: it can break toroidal symmetry and produce partial ergodisation of field lines further out in the pedestal gradient region, increasing overall particle and heat flux. Additionally, at a specific toroidal location in the gradient region (away from the current structure) a particular flux tube can escape through the broken separatrix (a homoclinic tangle, as described in [16]) and lead to the toroidally localised heat flux away from the main strike position shown in Fig. 6.

A long-lived vortex solution of ideal MHD is a non-linear effect driven by j×B and it is unlikely to be picked up with linear MHD simulation: it is not necessarily a linear instability. Nevertheless, study of the MHD stability of these plasmas is underway. We speculate that either the high $T_{e,ped}$ (ideal MHD) and/or the high rotation velocities present in the plasma before the mode onset leads to the circumstances necessary for localised current structures (MHD vortices) to form and survive. In [5] it is shown that localised solutions of the ideal MHD equations are described in term of the generalised velocities

$$\vec{w} = \left( \vec{v} \pm \frac{\vec{B}}{\sqrt{\mu_0 m_i n_i}} \right)$$

A condition for existence of a stationary magnetic vortex is that the growth of the generalised vorticity be zero. Based on [5] we suggest that such condition would be equivalent to the following relationship between toroidal velocity of plasma, background poloidal field and ion mass density:

$$\vec{w}_\perp \cdot \nabla \vec{w}_\perp^2 = 0, \text{ with } \vec{w}_\perp = \left( \vec{v}_{tor} \pm \frac{\vec{B}_{pol}}{\sqrt{\mu_0 m_i n_i}} \right)_\perp .$$

Here ⊥ implies perpendicular to the background magnetic field. This condition can produce both cyclonic and anti-cyclonic vortices (same sign or opposite sign of vorticity compared to background), possibly related to co and counter plasma rotation relative to plasma current direction. Another possible explanation of a localised rotating current structure is the bifurcation of the equilibrium field to a combined field produced by the background plasma plus a rotating helical perturbation, as sketched in [17]. More experiments are needed to test these theoretical predictions.

Interestingly, this plasma state as observed at JET during a long-lived OM is very reminiscent of the Quiescent H-mode [18, 19, 20], so named because it combines high confinement properties with the absence of ELMs. The localised current structure identified as the OM in JET may be related to the Edge Harmonic Oscillation (EHO) observed in quiescent H-modes of DIII-D and AUG. We suggest that a quasi-steady H-mode as induced by the OM in JET is a potentially useful operating regime, provided the basic physics governing the localised current structures can be sufficiently understood to warrant extrapolation to future fusion devices.

In summary: a spontaneously formed localised current structure has been observed in a JET tokamak plasma. It has been identified as a rotating helical ribbon located at the pedestal top. The current ribbon is long lived and has important effects on plasma behaviour: it increases transport across the high gradient region of the plasma pedestal and delays the appearance of ELMs. Theory-based predictions of localised structures in ideal MHD can be used to guide further research.

Acknowledgements:

We are grateful to Tom Osborne, W. Suttrop, and D. Borba for useful discussions, and to Paul Karman for the 3D representation of a current ribbon inside the JET vessel.

This work, supported by the European Communities under the contract of Association between EURATOM and the Laboratorio Nacional de Fusión, CIEMAT, was carried out within the framework of the European Fusion Development Agreement. The views and opinions expressed herein do not necessarily reflect those of the European Commission.

Figure Captions:

*Fig. 1: time traces showing a long-lived Outer Mode, starting from 14 s to 15.38s, and its effect on plasma behaviour: a) rise in $D_a$ ; b) rise in power to outer target; c) drop in pedestal temperatures; d) slower pedestal density rise, rising core density; e) stationary energy; f) toroidal rotation frequency at top and middle of pedestal g) FFT of magnetic signal*

*Fig. 2: Mirnov coil signals from toroidal coil array (outboard), showing toroidal propagation of magnetic feature in co-current direction. The lowest plot is the integrated signal from the coil at 3º*

*Fig. 3: Comparison of simulated signals(coloured) with data (black) in Mirnov coils located at the plasma outboard (top row), and three inboard coils located above, near and below the plasma midplane, respectively. Simulated signals are generated with rotating current ribbons located at q=4 (red), q=3 (blue) and q=2 (green) flux surfaces. The currents quoted in each case are adjusted to match calibrated signal amplitude.*

*Fig.4: q=4 current ribbon that provides best fit to magnetic signals from Mirnov coil set, with 100-200 A.*

*Fig. 5: profiles of toroidal rotation frequency, $n_e$, $T_e$, $T_i$ at (a) start and (b) end of OM*

*Fig. 6: heat flux contours at outer strike, measured with IR, and integrated magnetic signal showing correlation of heat pulses with rotation of current structure*

**References:**

[1] Birkeland, Kristian (1908), The Norwegian Aurora Polaris Expedition 1902-1903
[2] Walén, C., Ark. f. mat., astr. O. fysik, 30 A No. 15 and 31 B, No. 3
[3] Alfvén, H. Cosmical Electrodynamics, Oxford, England, Claredon Press, 1953
[4] Cummings, W.D., Dessler, J., Jour. Geophys. Res. **72**, p. 1007–1013, 1967
[5] Petviashvili, V.I., Plasma Phys. Rep. 19, (1993)
[6] Antar, G.Y. et al., *Nucl. Fusion* 49, 032001 (2009)
[7] Martines, E. et al., accepted for publication in PPCF.
[8] Nave, M.F.F. et al., Nucl. Fusion 35 409 (1995)
[9] Huysmans, G.T.A. et al., Nucl. Fusion 38, 179 (1998)
[10] Perez, C.P. et al., Nucl. Fusion 44 609-623 (2004)
[11] JET Team, presented by P.J. Lomas, Plas. Phys. & Contr. Nucl. Fus. Res. 1994 (Proc. 15th Int. Conf. Seville, 1994), Vol. 1, IAEA, Vienna 211 (1995).
[12] Smeulders, P. et al., Plasma Phys. Control. Fusion 41 1303 (1990)
[13] C P Perez et al., Plasma Phys. Control. Fusion 46 61-87 (2004)
[14] Nave, M.F.F. et al., Nucl. Fusion 39 1567 (1999)
[15] Solano, E.R. et al., Nucl. Fusion 48 065005 (2008)
[16] Evans, T.E. et al., J. Phys.: Conf. Ser. 7 174 (2005)
[17] F E M da Silveira and R.M.O Galvão, Plasma Phys. Control. Fusion 49, L11-15, (2007)
[18] Burrell, K.H. et al., Phys. Plas., 12, 056121 (2005)
[19] Burrell, K.H. et al., Phys. Rev. Lett. **102**, 155003 (2009)
[20] Suttrop, W. et al., Nucl. Fus. 45, 721 (2005)